\documentclass[conference]{IEEEtran}
\IEEEoverridecommandlockouts

\IEEEsettopmargin{t}{0.75in}

\usepackage{cite}
\usepackage{amsmath,amssymb,amsfonts}
\usepackage{algorithmic}
\usepackage{graphicx}
\usepackage{multirow}
\usepackage{booktabs}
\usepackage{xcolor}
\def\BibTeX{{\rm B\kern-.05em{\sc i\kern-.025em b}\kern-.08em
    T\kern-.1667em\lower.7ex\hbox{E}\kern-.125emX}}
\begin{document}

\title{A Study on Semi-Supervised Detection of DDoS Attacks under Class Imbalance
\thanks{This research was financially supported by the Natural Sciences and 
Engineering Research Council of Canada (NSERC).}
}

\author{\IEEEauthorblockN{Ehsan Hallaji}
\IEEEauthorblockA{\textit{Department of ECE} \\
\textit{University of Windsor}\\
Windsor ON, Canada \\
hallaji@uwindsor.ca}
\and
\IEEEauthorblockN{Vaishnavi Shanmugam}
\IEEEauthorblockA{\textit{Faculty of Computer Science} \\
\textit{University of New Brunswick}\\
Fredericton NB, Canada \\
v.s@unb.ca}
\and
\IEEEauthorblockN{Roozbeh Razavi-Far}
\IEEEauthorblockA{\textit{Faculty of Computer Science} \\
\textit{University of New Brunswick}\\
Fredericton NB, Canada \\
roozbeh.razavi-far@unb.ca}
\and
\IEEEauthorblockN{Mehrdad Saif}
\IEEEauthorblockA{\textit{Department of ECE} \\
\textit{University of Windsor}\\
Windsor ON, Canada \\
msaif@uwindsor.ca}
}

\maketitle

\begin{abstract}
One of the most difficult challenges in cybersecurity is eliminating Distributed Denial of Service (DDoS) attacks. Automating this task using artificial intelligence is a complex process due to the inherent class imbalance and lack of sufficient labeled samples of real-world datasets. This research investigates the use of Semi-Supervised Learning (SSL) techniques to improve DDoS attack detection when data is imbalanced and partially labeled. In this process, 13 state-of-the-art SSL algorithms are evaluated for detecting DDoS attacks in several scenarios. We evaluate their practical efficacy and shortcomings, including the extent to which they work in extreme environments. The results will offer insight into designing intelligent Intrusion Detection Systems (IDSs) that are robust against class imbalance and handle partially labeled data.
\end{abstract}


\section{Introduction}

DDoS is one of the most common cyber threats that make online services unavailable by overwhelming them with excessive traffic \cite{8735686, LI2023109895}. These attacks can result in significant operational and financial losses for organizations and enterprises. Moreover, DDoS attacks are often used as an initial step to more severe cyber attacks such as data breaches or network compromises. As DDoS attacks become more sophisticated, they are more likely to bypass the IDS. Thus, it is of paramount importance to maintain data security and digital trust in modern systems.

One of the recent advancements in this domain is the use of machine learning to build intelligent IDSs to detect cyber attacks \cite{LI2023109895, app14198840, 9993704, HALLAJI2023110384}. Using this approach, the IDS can process large volumes of data and quickly adapt to distribution changes. In addition, their ability to extract abstract features from the data facilitates analysis of dormant patterns that will be neglected by rule-based or expertise-oriented detection of DDoS attacks. Nevertheless, the data-driven nature of such systems necessitates using high-quality data for training the detection model.

Label scarcity and class imbalance are among the main challenges in designing intelligent IDS \cite{9563211, 9928313}. The labeling process is expensive and time-consuming and has to be repeated periodically to cope with evolving threats. SSL mitigates this problem by using a small number of labeled samples in combination with abundant unlabeled data \cite{lee2013pseudo,8401530, 8417973}. Nonetheless, DDoS datasets are often highly imbalanced \cite{8888419, electronics14010069}, as the majority of captured data is benign, and only a minority group corresponds to DDoS attacks. This will in turn cause a bias towards the benign samples in the model, raising the missed alarm rate. While Class Imbalance Learning (CIL) can potentially address this issue, their combination with SSL in a partially labeled setup is complex, as this combination may result in information loss or low-quality samples.

To tackle the aforementioned challenges, this work investigates the interaction of SSL and CIL for insights into how to build robust IDS that can effectively handle label scarcity and class imbalance. 
To do so, we experimentally review and evaluate 13 state-of-the-art SSL algorithms for detecting DDoS attacks. In this process, we study several ratios of class imbalance in combination with various labeling ratios to determine the robustness under various scenarios. 

The remainder of this paper is organized as follows. Section \ref{sec:problem} explains the targeted problem. Section \ref{sec:design} elaborates on the designed IDS and the employed case study. Section \ref{sec:results} reports and analyzes the experimental results. Finally, the paper is concluded in Section \ref{sec:conclusion}.

\section{Problem Statement}
\label{sec:problem}
One of the main challenges in training intelligent IDS is the process of labeling data in the preparation step. Collecting raw unlabeled data is often feasible in any application. However, categorizing them into certain categories and tagging them with a certain label is a time-consuming and often expensive task, as it requires a large volume of data to be reviewed by a team of experts. Given that the detection model should be updated periodically to keep up with the most recent attacks, it is crucial to obtain labeled data in relatively short intervals. To address the scarcity of labeled data, the literature resorts to SSL to enable the processing of partially labeled data during training. By doing so, only a very small portion of the data needs to be labeled, and the rest of the unlabeled data will be processed based on the distributional relationships with the labeled samples. 

Another issue of concern is the fact that the majority of the collected data samples are associated with benign conditions, and only a very small portion of data represents different classes of DDoS attacks. This imbalance in the data results in a bias in the detection model. In other words, the model learns the distribution of normal samples very well but misses the patterns associated with DDoS attacks, which in turn increases the missed alarm rate. Class imbalance can be eliminated using CIL approaches such as under-sampling and over-sampling. The former reduces the population of the majority class to that of the minority group, and the latter generates artificial samples replicating the distribution of the minority group to match the number of majority samples. 

Nonetheless, eliminating class imbalance in partially labeled data is a challenging task \cite{pmlr-v162-guo22e, 9563211}. Since the number of labeled samples is limited for both minority and majority groups, using under-sampling may result in significant information loss in the majority group, and over-sampling may produce bad samples that are not good representatives of the original distribution. To reach optimal performance, one should ideally find a trade-off between balancing the populations and maintaining the performance by minimizing information loss and production of redundant samples. In other words, the imbalance ratio will be reduced by CIL but not eliminated. Thus, SSL techniques still have to process imbalance data, albeit at a considerably lower imbalance ratio. This means that the SLL techniques should also be partially resilient against class imbalance to reach optimal detection performance.

\section{Simulation Setting}
\label{sec:design}
Here, simulation scenarios are first detailed, and employed detection models are introduced afterward.

\subsection{Scenarios}
To simulate DDoS attacks in the experiments, we employ CIC-DDoS2019 dataset \cite{8888419} and generate four imbalance ratios of 1:10, 1:100, 1:500, and 1:1000, indicating the number of majority samples for a single minority sample. For each imbalance ratio, 10 percent and 90 percent of samples are used for testing and training, respectively. Then, to make the training sets partially labeled, we randomly remove labels to obtain 1\%, 5\%, 10\%, and 20\% labeling ratios (i.e., number of labeled samples over all samples). The labeling ratios are applied to each class separately to ensure the ratio of labeled samples to the class population is consistent across all classes. In total, the combination of different imbalance and labeling ratios results in 16 different experimental scenarios in our simulations.

\subsection{Detection Models}
To distinguish DDoS attacks from benign samples, several state-of-the-art SSL classifiers are employed in our simulations as follows: AdaMatch \cite{berthelot2022adamatch}, FixMatch \cite{NEURIPS2020_06964dce}, FlexMatch \cite{NEURIPS2021_995693c1}, Label Propagation (LP) \cite{Iscen_2019_CVPR}, Label Spreading (LS)\cite{NIPS2003_87682805}, Mean Teacher (MT) \cite{NIPS2017_68053af2}, MixMatch \cite{NEURIPS2019_1cd138d0}, $\Pi$-Model \cite{laine2017temporal}, Pseudo-Labeling (LP) \cite{lee2013pseudo}, Self-Training (ST) \cite{stanescu2014semi}, Virtual Adversarial Training (VAT) \cite{8417973}, SimCLR \cite{pmlr-v119-chen20j}, and Suppressed Consistency Loss (SCL) \cite{pmlr-v162-guo22e}.

\section{Experimental Results}
\label{sec:results}
\begin{figure*}[t]
\includegraphics[trim={5.1cm 0.8cm 6.1cm 0.9cm},clip,width=\textwidth]{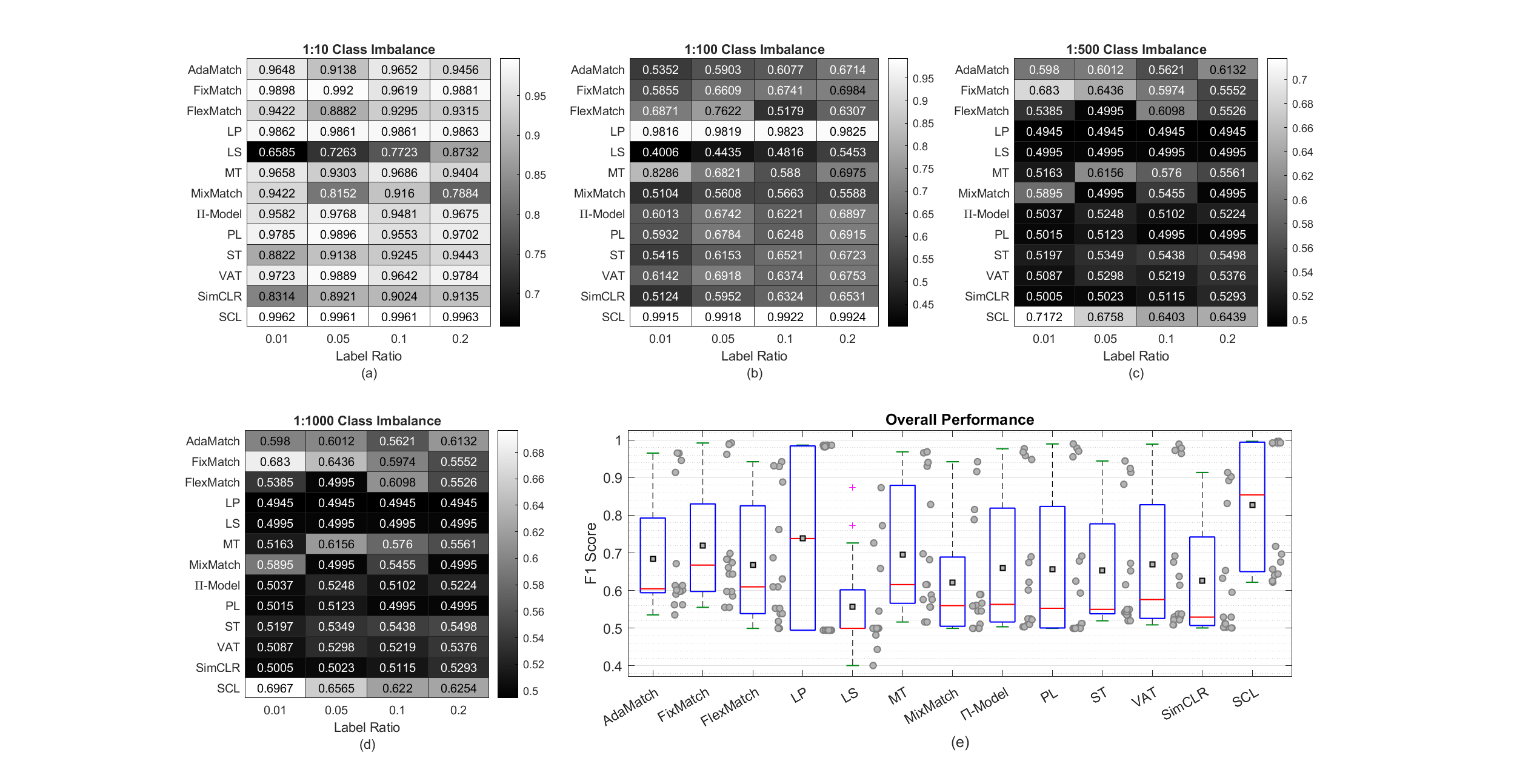}
\caption{Averaged F1-scores obtained through cross-validation. Heat maps show the performance obtained for certain degrees of class imbalance whereas the box plot illustrates the overall performance considering all imbalance and label ratios.}
\label{fig:plots}
\end{figure*}

This section first analyzes the obtained results based on the imbalance ratio used in the experiments in different subsections. These results are shown in Fig. \ref{fig:plots}. Then, the overall results are studied in the last subsection. All the experiments underwent 10-fold cross-validation and the associated standard deviation is reported accordingly.

\subsection{1:10 Imbalance Ratio}
The results associated with the 1:10 imbalance ratio in Fig. \ref{fig:plots}(a) show that SCL generally outperforms other methods in terms of F1 score. In addition, FixMatch demonstrates remarkable stability, peaking at 5\% labeled data with an F1 score of 0.9920 and maintaining strong performance across all ratios. Similarly, LP and ST excel by achieving F1 scores above 0.996, regardless of the label ratio. This showcases their robustness against small label ratios, making them reliable for various SSL scenarios.

Conversely, some algorithms exhibit variability and sensitivity to the label ratio. For instance, MixMatch shows a noticeable decline in performance as the label ratio increases. At 1\% labeled data, it achieves an F1 score of 0.9422, but it drops significantly to 0.7884 at 20\%. Similarly, FlexMatch struggles at 5\% labeled data (0.8882) but recovers somewhat at 10\% (0.9295) and 20\% (0.9315), though the improvement remains marginal. Additionally, LS shows slow and steady improvement, starting with a low F1 score of 0.6585 at 1\% and peaking at 0.8732 at 20\%. This slower adaptability highlights that some algorithms are less efficient in utilizing additional labeled data and may require further refinement.

VAT, UDA, and MT demonstrate strong adaptability, particularly with minimal labeled data. For instance, UDA achieves an impressive F1 score of 0.9740 at 1\% labeled data, making it one of the best-performing algorithms in low-data scenarios. VAT and MT also deliver reliable results, with MT peaking at 0.9686 at 10\% labeled data. These algorithms balance scalability and adaptability effectively, making them suitable for a wide range of applications where labeled data availability may vary. Their ability to maintain performance across different levels of labeled data sets them apart as robust solutions.

These results underscore that increasing the percentage of labeled data does not guarantee improved performance for all algorithms. While techniques such as FixMatch and ST benefit consistently from additional labeled data, others, such as MixMatch and FlexMatch, struggle to effectively balance labeled and unlabeled data as the percentage grows. Factors such as generalization ability, sensitivity to label noise, and the risk of overfitting play crucial roles in determining algorithm performance. Evaluating each algorithm’s adaptability and scalability under different levels of labeled data is essential for selecting the most suitable approach for a given task.

\subsection{1:100 Imbalance Ratio}

The results of the 1:100 ratio, illustrated in Fig. \ref{fig:plots}(b), highlight significant differences in performance compared to the 1:10 results, especially in their ability to leverage labeled data. Aside from SCL, most algorithms that excel in the 1:10 results struggle to maintain similar performance levels when scaled to the 1:100 ratio, reflecting sensitivity to dataset characteristics or limitations in generalization.

In the 1:100 results, FixMatch, which was a standout performer in the 1:10 results with a peak performance of 0.9920, shows a significant decline. Its performance at 1\% labeled data drops to 0.6830, and further decreases are observed as the labeled data percentage increases (e.g., 0.5552 at 20\%). This suggests that FixMatch struggles with the increased label sparsity and may over-rely on a small subset of labeled data in this scenario. Similarly, AdaMatch, which performed robustly in 1:10 with F1 scores over 0.960, achieves more modest results in 1:100, peaking at 0.6132 at 20\%.

LP, LS, and ST, which were consistent performers in the 1:10 results, fail to adapt in the 1:100 scenario. These algorithms maintain uniform F1 scores of 0.4995 across all label ratios, indicating that they cannot effectively utilize the sparse labeled data or integrate it with the unlabeled portion. This stark contrast underscores the limitations of these methods in handling more challenging data distributions or extreme label scarcity.

Interestingly, UDA, which excelled in 1:10 with F1 scores as high as 0.9740, continues to perform reasonably well at 1\% labeled data (0.7629) but drops significantly at higher label ratios. This suggests that UDA's effectiveness may be more dependent on the quality of unlabeled data than the absolute percentage of labeled data. Meanwhile, VAT, which demonstrated robustness in 1:10, remains relatively strong in 1:100, peaking at 0.6694 ± 0.1962 at 20\% labeled data. This indicates that VAT retains a better balance between leveraging labeled and unlabeled data compared to many other algorithms.

These comparisons reveal that increasing the labeled data ratio from 1:10 to 1:100 introduces significant challenges for most algorithms. Many methods, particularly those relying heavily on labeled data such as FixMatch and AdaMatch, struggle to maintain their effectiveness, whereas algorithms such as VAT and UDA show comparatively better scalability. These findings emphasize the importance of evaluating algorithms across different labeled data scenarios to ensure robustness and adaptability to varying data distributions.

\subsection{1:500 Imbalance Ratio}

Fig. \ref{fig:plots}(c) illustrates the results for the 1:500 ratio and shows further decline in performance for most algorithms compared to their performance in the 1:10 and 1:100 ratios, underscoring the challenges of extreme label sparsity. While the 1:100 ratio already exhibited reduced performance for many algorithms, the 1:500 ratio exacerbates this trend, highlighting limitations in scalability and adaptability. SCL that preserved its performance for previous imbalance rations also starts to lose performance with the 1:500 imbalance ration. However, it still outperforms the rest of the competitors in terms of F1 score.

FixMatch, which achieved a high performance in the 1:10 ratio with an F1 score of 0.9920 at 5\% labeled data, shows a steady decline in the 1:100 ratio (peaking at 0.6436) and deteriorates further in the 1:500 ratio (e.g., 0.5552 at 20\%). This steady drop suggests that FixMatch struggles to generalize as label imbalance increases. Similarly, AdaMatch, which performed robustly in the 1:10 ratio (e.g., 0.9652 at 10\% labeled data), performs modestly in the 1:100 ratio, peaking at 0.6132 at 20\%, and stagnates further in the 1:500 ratio, failing to surpass 0.6012 even at 5\%. These declines indicate that FixMatch and AdaMatch heavily rely on adequate labeled data and struggle in extreme imbalance scenarios.

UDA, another strong performer in the 1:10 ratio (e.g., 0.9740 at 1\%), also shows a notable decline as the imbalance increases. Its performance drops in the 1:100 ratio (e.g., 0.7629 at 1\%) and further in the 1:500 ratio, peaking at only 0.5787 at 20\%. In contrast, VAT demonstrates remarkable resilience across all three ratios. In the 1:10 ratio, VAT peaks at 0.9903 at 1\%, while in the 1:100 ratio, it maintains strong performance, achieving its best F1 score of 0.6694 at 20\%. Even in the 1:500 ratio, VAT achieves the same peak score of 0.6694 at 20\%, showing that its ability to balance labeled and unlabeled data remains robust despite increasing imbalance. This consistent performance across ratios highlights VAT’s adaptability to challenging scenarios and its ability to handle sparse labels better than most other algorithms.

Meanwhile, algorithms such as LP, LS, and ST fail to adapt in both the 1:100 and 1:500 ratios, maintaining uniform F1 scores of 0.4995 across all label ratios. This inability to leverage sparse labeled data underscores their limitations in highly imbalanced scenarios. Overall, while many algorithms such as FixMatch and UDA show significant performance declines as the imbalance increases, VAT stands out as the most resilient algorithm across all three ratios, consistently maintaining strong performance and proving to be a reliable choice for handling extreme label scarcity.

\subsection{1:1000 Imbalance Ratio}

As shown in Fig. \ref{fig:plots}(d), the results of 1:1000 ratio demonstrate a sharp decline in performance for most algorithms compared to the 1:100 and 1:500 ratios, highlighting the challenges of extreme label sparsity. While some algorithms retained modest performance in the 1:100 and 1:500 ratios, the 1:1000 ratio sees most methods stagnating around baseline scores, often failing to leverage the labeled data effectively. This indicates fundamental limitations in scalability and adaptability under extreme imbalance scenarios.

Similar to the previous analysis, SCL outperforms all methods with this ratio as well. FixMatch, which performed reasonably well in the 1:100 ratio with a peak F1 score of 0.6436 at 5\%, and maintained a score of 0.5552 at 20\% in the 1:500 ratio, experiences a drastic decline in the 1:1000 ratio, peaking only at 0.5466 at 5\%. At other label ratios, its performance stagnates at 0.4998, indicating its inability to adapt to the extreme imbalance. Similarly, AdaMatch, which achieved 0.6012 at 5\% in the 1:500 ratio, completely fails in the 1:1000 ratio, with all scores flatlining at 0.4998.

UDA, a strong performer in the 1:100 ratio (e.g., 0.7629 at 1\%), also shows significant decline by the 1:500 ratio (e.g., 0.5787 at 20\%), and fares even worse in the 1:1000 ratio, peaking at 0.5606 at 5\%. This steady deterioration suggests that UDA's reliance on unlabeled data becomes less effective as labeled data becomes sparser. VAT, one of the most resilient algorithms in the 1:100 and 1:500 ratios, also suffers in the 1:1000 ratio. Its peak F1 score of 0.6694 at 20\% in the 1:100 and 1:500 ratios drops to 0.5475 at 1\%, with similar scores across other percentages (e.g., 0.5449 at 20\%). This decline reflects the challenges of extreme label scarcity, likely limiting VAT’s ability to establish robust decision boundaries using adversarial perturbations. However, VAT still performs better than many other algorithms, demonstrating relative stability in its scores.

Interestingly, MT and MixMatch show some adaptability in the 1:1000 ratio, with MT peaking at 0.6402 at 10\% and MixMatch reaching 0.5590 at 10\%. While their scores are lower than in the 1:100 and 1:500 ratios, they indicate slightly better scalability compared to most other methods. PL also outperforms several algorithms, achieving 0.5940 at 1\%, though its performance declines at higher label ratios.

In contrast, algorithms such as LP, LS, and ST, which failed in both the 1:100 and 1:500 ratios, remain completely ineffective in the 1:1000 ratio, with uniform scores of 0.4995 across all label ratios.

\subsection{Overall Results}
Fig. \ref{fig:plots}(e) illustrates the overall results across all scenarios. Imbalance in labeled data significantly affects the performance of SSL algorithms, as observed across the 1:10, 1:100, 1:500, and 1:1000 ratios. With increasing imbalance, most algorithms show a sharp decline in their F1 scores, reflecting their limited ability to adapt to sparse labeled data. In the 1:10 ratio, the average F1 score is approximately 0.960, with strong performances from algorithms such as SCL, FixMatch, VAT, and UDA, all peaking above 0.97. This highlights how sufficient labeled data enables most algorithms to achieve near-optimal performance. However, as the imbalance increases, the average F1 score drops significantly, with many algorithms struggling to maintain their effectiveness.

In the 1:100 ratio, the average F1 score declines to around 0.670, indicating the challenges posed by reduced labeled data. SCL, VAT and UDA continue to perform reasonably well, with VAT peaking at 0.6694 at 20\%, showcasing its ability to balance labeled and unlabeled data effectively. However, many algorithms, such as LP and ST, fail to adapt, flatlining at baseline F1 scores of 0.4995 across all label ratios. In the 1:500 ratio, the average F1 score drops further to approximately 0.550, with only a few algorithms such as VAT maintaining relative resilience. VAT retains its peak score of 0.6694, while algorithms such as FixMatch and AdaMatch show sharp declines, reflecting their inability to handle extreme imbalance.

The 1:1000 ratio presents the most extreme scenario, with the average F1 score dropping sharply to around 0.510. Most algorithms stagnate near baseline scores, failing to leverage the limited labeled data effectively. SCL, while declining compared to previous ratios, remains one of the better-performing algorithms. Similarly, MT and PL exhibit some adaptability, with MT achieving 0.6402 at 10\% and PL reaching 0.5940 at 1\%. These performances, though modest, are higher than those of algorithms such as FixMatch, which flatline at 0.4998, indicating their inability to scale under extreme imbalance.

Despite its relative robustness, even SCL shows limitations as the imbalance grows. While it maintains stability across the 1:10, 1:100, and 1:500 ratios, its decline in the 1:1000 ratio highlights the challenges of establishing robust decision boundaries with adversarial perturbations when labeled data is scarce. Algorithms such as LP, LS, and ST fail under imbalance, consistently flatlining at baseline scores across all ratios beyond 1:10, underscoring their inability to leverage sparse labeled data or the geometric structure of unlabeled data.

The increasing imbalance from 1:10 to 1:1000 reveals the limitations of most algorithms in handling extreme label scarcity. While algorithms such as SCL, VAT, MT, and PL perform better than others, the sharp drop in average F1 scores highlights the need for methods specifically designed to address imbalance. The findings underscore the importance of developing robust SSL algorithms that can effectively utilize unlabeled data and adapt to varying levels of labeled data imbalance.

\section{Conclusion}
\label{sec:conclusion}
This paper performed an experimental review on 13 state-of-the-art SSL algorithms for detecting DDoS algorithms under harsh conditions such as a high imbalance ratio and scarcity of labels in data. These methods underwent 16 different simulation scenarios with various labeling and imbalance ratios on a real-world DDoS dataset. Results have demonstrated the efficiency of some SSL methods such as SCL in highly imbalanced scenarios and the inability of others to optimally use partially labeled data. These findings underscore the need for robust SSL techniques that balance scalability and adaptability to meet the challenges of modern cybersecurity threats.
\bibliographystyle{IEEEtran}
\bibliography{main}

\end{document}